\input{psfig}
\documentstyle[referee]{laa}
\thesaurus{12.03.4; 11.06.1}
\title{Tidal torques dynamical friction and the structure of clusters 
of galaxies\\ } 
\author{A. ~Del Popolo\inst{}, M. Gambera\inst{} }
\offprints{A. ~Del Popolo, M. ~Gambera}
\institute{Istituto di Astronomia dell'Universit\`a di Catania, \\
Citt\`a Universitaria, Viale A.Doria, 6 - I 95125 Catania, Italy}
\date{}
\begin{document}
\maketitle
\begin{abstract}
We study the joint effect of tidal torques and dynamical friction on the
collapse of density peaks solving numerically the equations of motion of a
shell of barionic matter falling into the central regions of a cluster of
galaxies. We calculate the evolution of the expansion parameter, $a(t)$, of
the perturbation using a coefficient of dynamical friction $\eta _{cl}$
obtained from a clustered system and taking into account the gravitational
interaction of the quadrupole moment of the system with the tidal field of
the matter of the neighboring proto-galaxies. We show that within
high-density environments, such as rich clusters of galaxies, tidal torques
and dynamical friction slow down the collapse of low-$\nu $ peaks producing
an observable variation of the parameter of expansion of the shell. As a
consequence a bias of dynamical nature arises because high-density peaks
preferentially collapse to form halos within which visible objects
eventually will condense. For a standard Cold Dark Matter model this ${\it %
dynamical}$ ${\it bias}$ can account for a substantial part of the total
bias required by observations on cluster scales.
\end{abstract}
\begin{flushleft}
{\bf 1.Introduction.}
\end{flushleft}
The origin and evolution of large scale structures is nowadays the
outstanding problem in Cosmology. In a hierarchical 'bottom-up' scenario,
high density collapsed peaks cluster and merge to forme larger structures. The
density fluctuation field is often assumed to be locally isotropic, the
amplitudes are Gaussian distributed with uncorrelated phases (Peebles 1980).
Under these assumptions the power spectrum for a continous random field is
basically the squared amplitude of its Fourier modes. If we assume, for
computational convenience, that the random field $\delta (r)$ is periodic in
some large rectangular volume V, we can define the Fourier trasform to be:
\begin{equation}
\delta ({\bf k})=\frac 1V\int \delta ({\bf r})\exp (i{\bf kr})d^3{\bf r} 
\end{equation}
According to Wiener-Khintchine theorem:
\begin{equation}
<|\delta _k|^2>=\frac 1V\int \xi ({\bf r})\exp (i{\bf kr})d^3 {\bf r} 
\end{equation}
and
\begin{equation}
\xi ({\bf r})=\frac V{2\pi ^3}\int <|\delta _k|^2>\exp (-i{\bf kr})d^3{\bf r}
\end{equation}
i.e. the power spectrum, $P(k)$, is the Fourier trasform of the
autocorrelation function and vice versa. Also the density fluctuation field, 
$\delta (r)$, can be obtained from the Fourier trasform of the power
spectrum, $P(k)$. In other words, on average the characteristics of the
density field peaks, e.g., their mass distribution, peculiar velocities,
etc., are completely determined by the spectrum through its moments (at
least during the linear and early non-linear phases of the collapse (Bardeen
et al. 1986)). Moreover the isotropy condition imposes that all physical
quantities around density peaks is, on average, spherically symmetric.
However, actual realizations of, e.g., the density field distributions
around the density peaks which eventually will give birth to galaxies and
clusters, depart from spherical simmetry and from the average density
profile, producing important consequences on collapse dynamics and formation
of protostructures (Hoffman \& Shaham 1985; Ryden 1988; Heavens \& Peacock
1988; Kashlinsky 1986, 1987; Peebles 1990). A fundamental role in this
context is played by the joint action of tidal torques (coupling shells of
matter which are accreted around a density peak and neighboring
protostructures (Ryden 1988)), and by dynamical friction (White 1976;
Kashlinsky, 1986, 1987, Antonuccio \& Colafrancesco 1995 (hereafter AC), 
Del Popolo \& Gambera 1996).
Some authors (see Barrow \& Silk 1981, Szalay
\& Silk 1983, Peebles 1990) have proposed that non-radial motions would
be expected within a developing proto-cluster, due to the tidal interaction
of the irregular mass distribution around them, (typical of hierarchical
clustering models), with the neighboring proto-clusters. 
The kinetic energy of this non-radial motions opposes the collapse of the
proto-cluster, enabling the same to reach statistical equilibrium before the
final collapse (the so called previrialization conjecture by Davis \&
Peebles 1977, Peebles 1990).
Non-radial motions change the
energetics of the collapse model 
by introducing another potential energy term.
One expects that non-radial motions produce firstly a change in the turn
around epoch, secondly a new functional form for $\delta _c$, thirdly 
a change of the mass function calculable with the Press-Schechter (1974)
formula and finally a modification of the two-point correlation function.
Recently Colafrancesco, Antonuccio \& Del Popolo (1995, hereafter CAD) 
have shown that dynamical friction delays the collapse of low-$\nu $ peaks
inducing a bias of dynamical nature. Because of dynamical friction
under-dense regions in clusters (the clusters outskirts) accrete less mass
with respect to that accreted in absence of this dissipative effect and as a
consequence over-dense regions are biased toward higher mass (Antonuccio \&
Colafrancesco 1995 and Del Popolo \& Gambera, 1996). Dynamical friction and 
non-radial motions acts
in a similar fashion: they delay the shell collapse
consequently inducing a dynamical bias similar to that produced by dynamical
friction but obviously of a larger value. 
This dynamical bias can be evaluated defining a selection function
similar to that given in CAD and using Bardeen, Bond, Szalay and Kaiser
(1986, hereafter BBKS) prescriptions. 

The plan of the paper is the following: in \S 2 we obtain the total specific
angular momentum acquired during expansion by a proto-cluster. In \S 3 we
calculate the dynamical friction force for galaxies moving into the cluster,
taking account of the clustering. In \S 4 we
use the calculated specific angular momentum and the dynamical friction force
to obtain the time of collapse
of shells of matter around peaks of density having $\nu _c=2,3,4$ and we
compare the results with Gunn \& Gott's (1972, hereafter GG) spherical
collapse model. In \S 5 we derive a selection function for the peaks giving
rise to proto-structures while in \S 6 we calculate some values for the bias
parameter, using the selection function derived, on three relevant filtering
scales. 
Finally in \S 7 we discuss the results
obtained.

{\bf 2. Tidal torques and angular momentum.}

The explanation of galaxies spins gain through tidal torques was pioneered
by Hoyle (1949). Peebles (1969) performed the first detailed calculation of the
acquisition of angular momentum in the early stages of protogalactic
evolution. More recent analytic computations (White 1984, Hoffman 1986,
Ryden 1988a) and numerical simulations (Barnes \& Efstathiou 1987) have
re-investigated the role of tidal torques in originating galaxies angular
momentum. \\
One way to study the variation of angular momentum with radius in
a galaxy is that followed by Ryden (1988a). In this approach the protogalaxy
is divided into a series of mass shells and the torque on each mass shell is
computed separately. The density profile of each proto-structure is
approximated by the superposition of a spherical profile, $\delta (r)$, and
a random CDM distribution, ${\bf \varepsilon (r)}$, which provides the
quadrupole moment of the protogalaxy. 
As shown by Ryden (1988a) the net rms torque on a
mass shell centered on the origin of internal radius $r$ and thickness $%
\delta r$ is given by: 
\begin{eqnarray}
\langle |\tau |^2\rangle ^{1/2}=\sqrt{30}\left( \frac{4\pi }5G\right) 
[\langle a_{2m}(r)^2\rangle \langle q_{2m}(r)^2\rangle \nonumber \\
-\langle a_{2m}(r)q_{2m}^{*}(r)\rangle ^2] ^{1/2}  \label{eq:tau}
\end{eqnarray}
where $q_{lm}$, the multipole moments of the shell and $a_{lm}$, the tidal
moments, are given by: 
\begin{equation}
\langle q_{2m}(r)^2\rangle =\frac{r^4}{\left( 2\pi \right) ^3}M_{sh}^2\int
k^2dkP\left( k\right) j_2\left( kr\right) ^2
\end{equation}
\begin{equation}
\langle a_{2m}(r)^2\rangle =\frac{2\rho _b^2r^{-2}}\pi \int dkP\left(
k\right) j_1\left( kr\right) ^2
\end{equation}
\begin{equation}
\langle a_{2m}(r)q_{2m}^{*}(r)\rangle =\frac r{2\pi ^2}\rho _bM_{sh}\int
kdkP\left( k\right) j_1\left( kr\right) j_2(kr)
\end{equation}
where $M_{sh}$ is the mass of the shell, $j_1(r)$ and $j_2(r)$ are the
spherical Bessel function of first and second order while the power spectrum 
$P(k)$ is given by: 
\begin{eqnarray}
P(k) &=&Ak^{-1}\left[ \ln \left( 1+4.164k\right) \right] ^2  \nonumber \\
&&\ \ ( 192.9+1340k+1.599\times 10^5k^2+1.78\times 10^5k^3+\nonumber \\
&&\ 3.995\times 10^6k^4)^{-1/2}
\end{eqnarray}
(Ryden \& Gunn 1987). The normalization constant $A$ can be obtained, as
usual, imposing that the mass variance at $8h^{-1}Mpc$, $\sigma _8$, is
equal to unity. 
Filtering the spectrum on cluster scales, $R_f=3h^{-1}Mpc$, we have obtained
the rms torque, $\tau (r)$, on a mass shell using Eq. (\ref{eq:tau}) then we
obtained the total specific angular momentum, $h(r,\nu )$, acquired during
expansion integrating the torque over time (Ryden 1988a Eq. 35): 
\begin{eqnarray}
h(r,\nu )=\frac 13\left( \frac 34\right) ^{2/3} \nonumber \\
\frac{\tau _ot_0}{M_{sh}}%
\overline{\delta }_o^{-5/2}\int_0^\pi \frac{\left( 1-\cos \theta \right) ^3}{%
\left( \vartheta -\sin \vartheta \right) ^{4/3}}\frac{f_2(\vartheta )}{%
f_1(\vartheta )-f_2(\vartheta )\frac{\delta _o}{\overline{\delta _o}}}%
d\vartheta   \label{eq:ang}
\end{eqnarray}
the functions $f_1(\vartheta )$, $f_2(\vartheta )$ are given by Ryden
(1988a) (Eq. 31) while the mean over-density inside the shell, $\overline{%
\delta }(r)$, is given by Ryden (1988a): 
\begin{equation}
\overline{\delta }(r,\nu )=\frac 3{r^3}\int_0^\infty d\sigma \sigma ^2\delta
(\sigma )
\end{equation}
\begin{figure}
\psfig{file=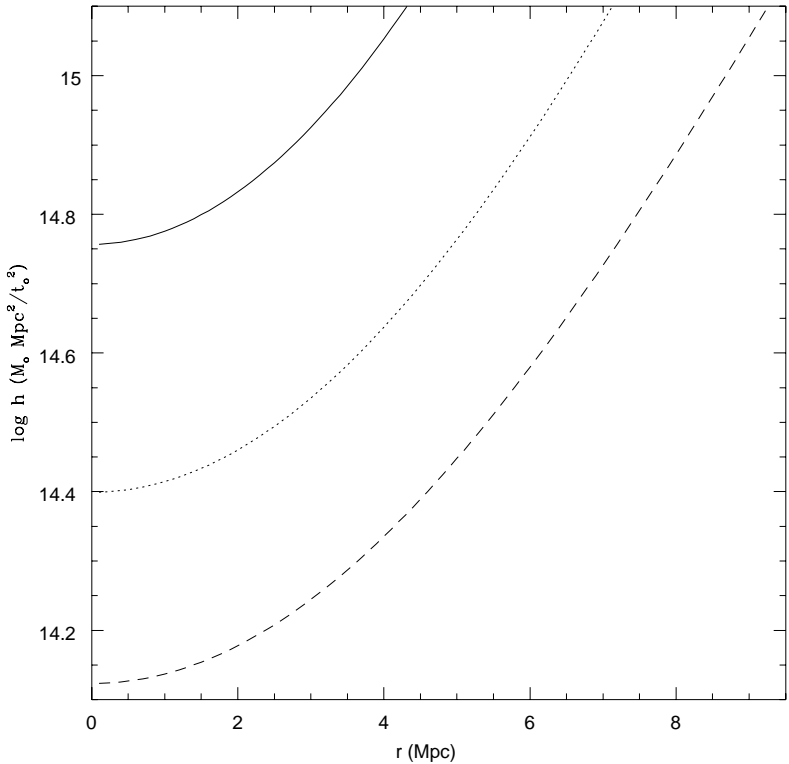,width=11cm,height=11cm}
\caption[]{The specific angular momentum, in units of $M_{\odot}$, Mpc and 
the Hubble time, $t_{o}$, for
three values of the parameter $\nu$ ($\nu=2$ solid line, $\nu=3$
dotted line, $\nu=4$ dashed line)  and for $R_{f}=3h^{-1}Mpc$.}
\end{figure}
In fig. 1 we show the variation of $h(r,\nu )$ with the distance $r$ for
three values of the peak height $\nu $. The rms specific angular momentum, $%
h(r,\nu )$, increases with distance $r$ while peaks of greater $\nu $
acquire less angular momentum via tidal torques. This is the angular
momentum-density anticorrelation showed by Hoffman (1986). This effect
arises because the angular momentum is proportional to the gain at turn
around time, $t_m$, which in turn is proportional to $\overline{\delta }%
(r,\nu )^{-\frac 32}\propto \nu ^{-3/2}$.

{\bf 3. Dynamical friction. } 

In a hierarchical structure formation model, the large scale cosmic
environment can be represented as a collisionless medium made of a hierarchy
of density fluctuations whose mass, $M$, is given by the mass function $%
N(M,z)$, where $z$ is the redshift. In these models matter is concentrated
in lumps, and the lumps into groups and so on.
In such a material system, gravitational
field can be decomposed into an average field, ${\bf F}_0(r)$, generated
from the smoothed out distribution of mass, and a stochastic component, $%
{\bf F}_{stoch}(r)$, generated from the fluctuations in number of the field
particles. 
The stochastic component of the gravitational field is specified assigning a
probability density, $W({\bf F})$, (Chandrasekhar \& von Neumann 1942). In
an infinite homogeneous unclustered system $W({\bf F})$ is given by
Holtsmark distribution (Chandrasekhar \& von Neumann 1942) while in
inhomogeneous and clustered systems $W({\bf F})$ is given by Kandrup (1980)
and Antonuccio-Delogu \& Barandela (1992) respectively. The stochastic
force, ${\bf F}_{stoch}$, in a self-gravitating system modifies the motion
of particles as it is done by a frictional force. In fact a particle moving
faster than its neighbours produces a deflection of their orbits in such a
way that average density is greater in the direction opposite to that of
traveling causing a slowing down in its motion. \\ Following Chandrasekhar
\& von Neumann's (1942) method, the frictional force which is experienced by
a body of mass $M$ (galaxy), moving through a homogeneous and isotropic
distribution of lighter particles of mass $m$ (substructure), having a
velocity distribution $n(v)$ is given by: 
\begin{equation}
M\frac{d{\bf v}}{dt}=-4\pi G^2M^2n(v)\frac{{\bf v}}{v^3}\log \Lambda \rho 
\label{eq:cha}
\end{equation}
where $\log \Lambda $ is the Coulomb logarithm, $\rho $ the density of the
field particles (substructure). \\ A more general formula is that given by
Kandrup(1980) in the hypothesis that there are no correlations among random
force and their derivatives: 
\begin{equation}
{\bf F}=-\eta {\bf v}=-\frac{\int W(F)F^2T(F)d^3F}{2<v^2>}{\bf v}
\end{equation}
where $\eta $ is the coefficient of dynamical friction, $T(F)$ the average
duration of a random force impulse, $<v^2>$ the characteristic speed of a
field particle having a distance $r\simeq (\frac{GM}F)^{1/2}$ from a test
particle (galaxy). This formula is more general than Eq. (\ref{eq:cha})
because the frictional force can be calculated also for inhomogeneous
systems when $W(F)$ is given. If the field particles are distributed
homogeneously the dynamical friction force is given by: 
\begin{equation}
F=-\eta v=-\frac{4.44G^2m_a^2n_a}{[<v^2>]^{3/2}}\log \left\{ 1.12\frac{<v^2>%
}{Gm_an_a^{1/3}}\right\} 
\end{equation}
(Kandrup 1980), where $m_a$ and $n_a$ are respectively the average mass and
density of the field particles. Using virial theorem we also have: 
\begin{equation}
\frac{<v^2>}{Gm_an_a^{1/3}}\simeq \frac{M_{tot}}m\frac
1{n^{1/3}R_{sys}}\simeq N^{2/3}
\end{equation}
where $M_{tot}$ is the total mass of the system, $R_{sys}$ its radius and $N$
is the total number of field particles. The dynamical friction force can be
written as follows: 
\begin{equation}
F=-\eta v=-\frac{4.44[Gm_an_{ac}]^{1/2}}N\log \left\{ 1.12N^{2/3}\right\}
\frac v{a^{3/2}}
\end{equation}
where $N=\frac{4\pi }3R_{sys}^3n_a$ and $n_{ac}=n_a\times a^3$ is the
comoving number density of peaks of substructure of field particles. This
last equation supposes that the field particles generating the stochastic
field are virialized. This is justified by the previrialization hypothesis
(Davis \& Peebles 1977). \\ To calculate the dynamical evolution of the
galactic component of the cluster it is necessary to calculate the number
and average mass of the field particles generating the stochastic field. \\ %
The protocluster, before the ultimate collapse at $z\simeq 0.02$, is made of
substructure having masses ranging from $10^6-10^9M_{\odot }$ and from
galaxies. We suppose that the stochastic gravitational field is generated
from that portion of substructure having a central height $\nu $ larger than
a critical threshold $\nu _c$. This latter quantity can be calculated
(following AC) using the condition that the peak radius, $r_{pk}(\nu \ge \nu
_c),$ is much less than the average peak separation $n_a(\nu \ge \nu
_c)^{-1/3}$, where $n_a$ is given by the formula of BBKS for the upcrossing
points: 
\begin{eqnarray}
n_{ac}(\nu \ge \nu _c)=\frac{\exp (\nu _c^2/2)}{(2\pi )^2}(\frac \gamma
{R_{*}})^3 [ \nu _c^2-1+ \nonumber \\
\frac{4\sqrt{3}}{5\gamma ^2(1-5\gamma ^2/9)^{1/2}}
\exp ( -5\gamma ^2\nu _c^2/18)] 
\end{eqnarray}
where $\gamma $, $R_{*}$ are parameters related to moments of the power
spectrum (BBKS Eq. ~4.6A). The condition $r_{pk}(\nu \ge \nu _c)<0.1n_a(\nu
\ge \nu _c)^{-1/3}$ ensures that the peaks of substructure are point like.
Using the radius for a peak: 
\begin{equation}
r_{pk}=\sqrt{2}R_{*}\left[ \frac 1{(1+\nu \sigma _0)(\gamma ^3+(0.9/\nu
))^{3/2}}\right] ^{1/3}
\end{equation}
(AC), we obtain a value of $\nu _c=1.3$ and then we have $n_a(\nu \ge \nu
_c)=50.7Mpc^{-3}$ 
($\gamma =0.4$, $R_{*}=50Kpc$) and $m_a$ is given by: 
\begin{equation}
m_a=\frac 1{n_a(\nu \ge \nu _c)}\int_{\nu _c}^\infty m_{pk}(\nu )N_{pk}(\nu
)d\nu =10^9M_{\odot }
\end{equation}
(in accordance with the result of AC), where $m_{pk}$ is given in Peacock $\&
$ Heavens (1990) and $N_{pk}$ is the average number density of peak (BBKS
Eq. ~4.4).
Clusters of galaxies are correlated systems whose autocorrelation function
can be expressed, in the range $10h^{-1} Mpc<r<60 h^{-1} Mpc$,
in a power law form:
\begin{equation}
\xi_{cc}=(\frac{r_{o,c}}{r})^{\gamma}
\end{equation}
with $\gamma \simeq 1.8$ and a correlation length,
$r_{o,c} \simeq 25h^{-1} Mpc$ (Bahcal \& Soneira 1983; Postman et al. 1986).
The analysis of fair samples of galaxies gives for the galaxy autocorrelation
function the expression:
\begin{equation}
\xi_{gg}=(\frac{r_{o,g}}{r})^{\gamma}
\end{equation}
in the range $0.1h^{-1} Mpc<r<10 h^{-1} Mpc$ ($r_{o,g} \simeq 5 h^{-1} Mpc$,
$\gamma=1.77 \pm 0.03$ (Peebles 1980, Davis \& Peebles 1983)). 
The description of dynamical friction in these systems need to
use a distribution of the stochastic forces, $W(F)$, taking account of
correlations. In this last case the coefficient of dynamical friction,
$ \eta$, may be calculated using the equation:
\begin{equation}
\eta=\int  d^{3} {\bf F} W(F) F^{2} T(F)/(2<v^{2}>)
\end{equation}
and using Antonuccio \& Atrio (1992) distribution:
\begin{equation}
W(F)=\frac{1}{2 \pi^2 F} \int_{0}^{\infty} dk k sin(kF)A_{f}(k)
\end{equation}
where $A_{f}$ is given in the quoted paper. The function $A_{f}$
is a linear integral function of the correlation function $ \xi(r)$.
As shown in Del Popolo \& Gambera (1997) the effect of clustering is
that to increase the effects of dynamical friction.

{\bf 4. Shell collapse time.}
Tidal torques and dynamical friction acts in a similar fashion.  
As known one of the consequences of the angular momentum acquisition 
by a mass shell
\begin{figure}
\psfig{file=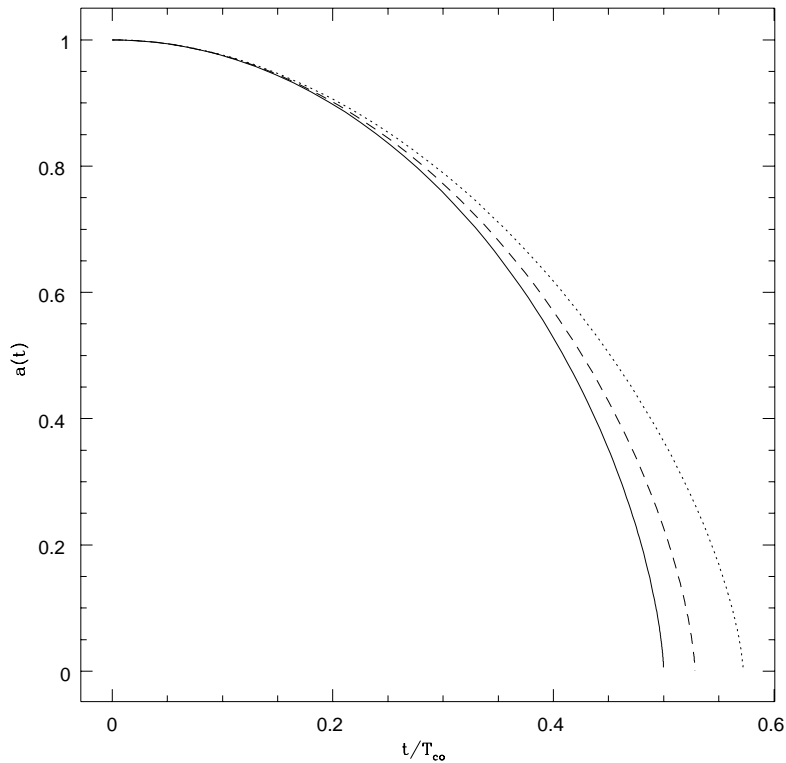,width=11cm,height=11cm}
\caption[]{The time evolution of the expansion parameter. The solid 
line is a(t) for GG model; the dashed line is a(t) taking account only
dynamicalfriction; the dotted line is a(t) taking account of the cumulative
effect of non-radial motions and dynamical friction.}
\end{figure}
of a proto-cluster is the delay of the collapse of the proto-structure. As
shown by Barrow \& Silk (1981) and Szalay \& Silk (1983) the gravitational
interaction of the irregular mass distribution of proto-cluster with the
neighbouring proto-structures gives rise to non-radial motions, within the
protocluster, which are expected to slow the rate of growth of the density
contrast and to delay or suppress collapse. According to Davis \& Peebles
(1977) the kinetic energy of the resulting non-radial motions at the epoch
of maximum expansion increases so much to oppose the recollapse of the
proto-structure. Numerical N-body simulations by Villumsen \& Davis (1986)
showed a tendency to reproduce this so called previrialization effect. In a
more recent paper by Peebles (1990) the slowing of the growth of density
fluctuations and the collapse suppression after the epoch of the maximum
expansion were re-obtained using a numerical action method. 
In the central regions of a density peak ($r\leq 0.5R_f$) the velocity
dispersion attain nearly the same value (Antonuccio \& Colafrancesco 1997)
while at larger radii ($r\geq R_f$) the radial component is lower than the
tangential component. This means that motions in the outer regions are
predominantly non-radial and in these regions the fate of the infalling
material could be influenced by the amount of tangential velocity relative
to the radial one. This can be shown writing the equation of motion of a
spherically symmetric mass distribution with density $n(r)$ (Peebles 1993): 
\begin{equation}
\frac \partial {\partial t}n\langle v_r\rangle +\frac \partial {\partial
r}n\langle v_r^2\rangle +\left( 2\langle v_r^2\rangle -\langle v_\vartheta
^2\rangle \right) \frac nr+n(r)\frac \partial {\partial t}\langle v_r\rangle
=0  \label{eq:peeb}
\end{equation}
where $\langle v_r\rangle $ and $\langle v_\vartheta \rangle $ are,
respectively, the mean radial and tangential streaming velocity. Eq. (\ref
{eq:peeb}) shows that high tangential velocity dispersion $(\langle
v_\vartheta ^2\rangle \geq 2\langle v_r^2\rangle )$ may alter the infall
\begin{figure}
\psfig{file=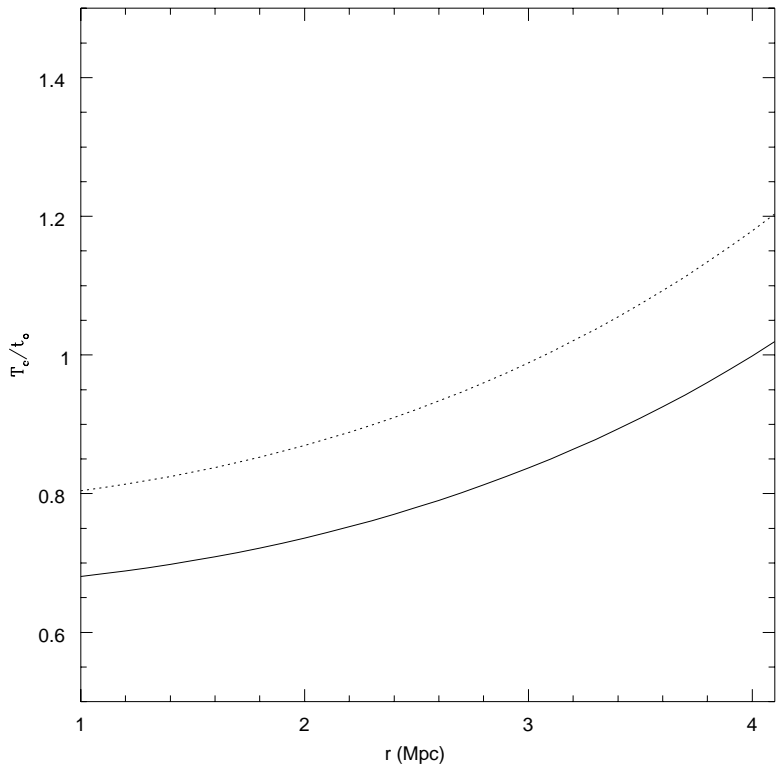,width=11cm,height=11cm}
\caption[]{The time of collapse of a shell of matter in units of the age of the
universe $t_{o}$ for $\nu=2$ (dotted line) compared with Gunn \& Gott's
model (solid line).}
\end{figure}
pattern. The expected delay in the collapse of a perturbation, due to
non-radial motions and dynamical friction, may be
calculated solving the equation for the radial acceleration (Kashlinsky
1986, 1987; AC; Peebles 1993): 
\begin{equation}
\frac{dv_r}{dt}=\frac{L^2(r,\nu )}{M^2r^3}-g(r) -\eta \frac{dr}{dt} \label{eq:coll}
\end{equation}
\begin{figure}
\psfig{file=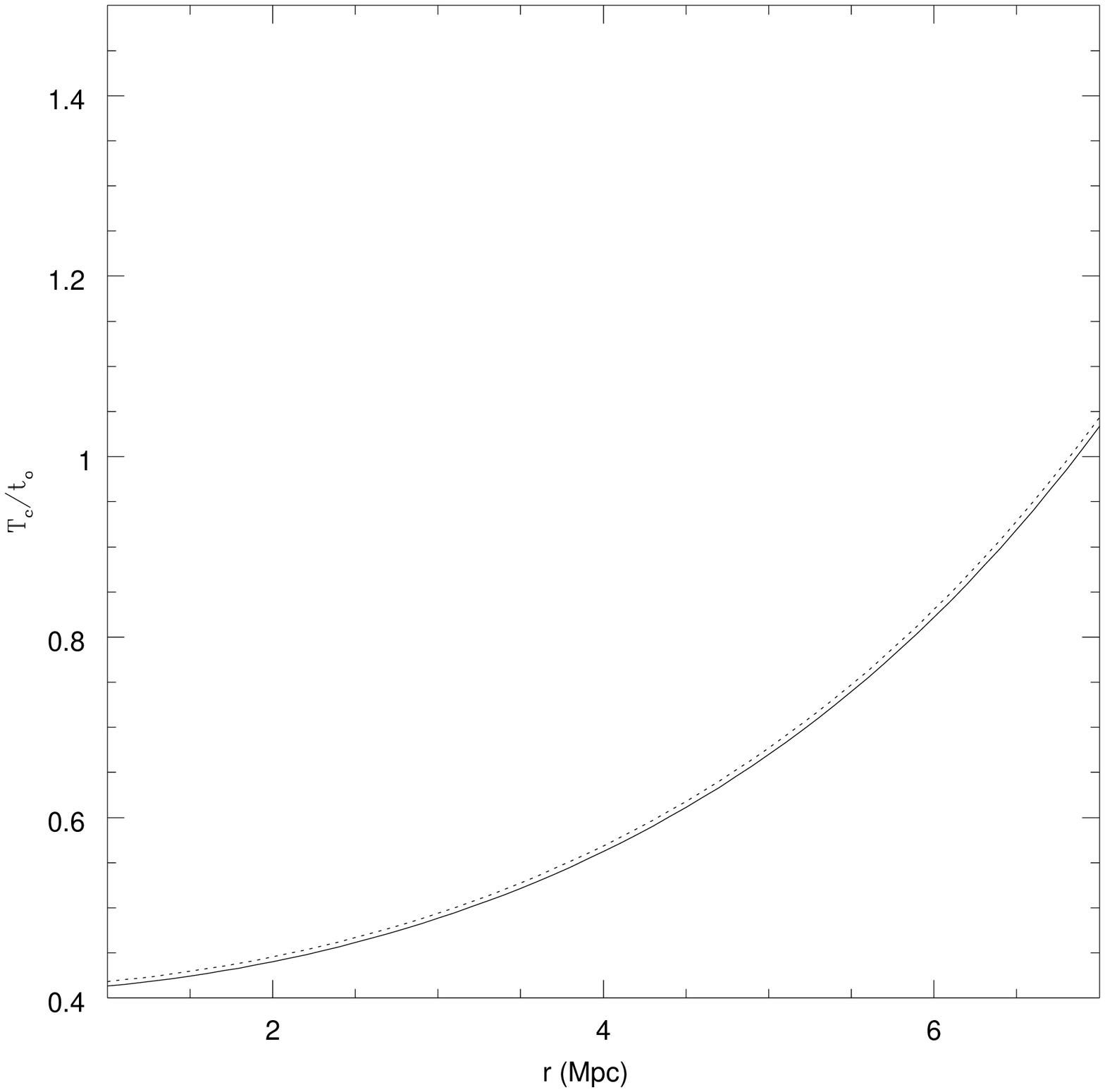,width=11cm,height=11cm}
\caption[]{The time of collapse of a shell of matter in units of the age of the
universe $t_{o}$ for $\nu=3$ (dotted line) compared with Gunn \& Gott's
model (solid line).}
\end{figure}
where $L(r,\nu )$ is the angular momentum and $g(r)$ the acceleration.
Writing the proper radius of a shell in terms of the expansion parameter, $%
a(r_i,t)$: 
\begin{equation}
r(r_i,t)=r_ia(r_i,t)
\end{equation}
remembering that 
\begin{equation}
M=\frac{4\pi }3\rho _b(r_i,t)a^3(r_i,t)r_i^3
\end{equation}
and that $\rho _b=\frac{3H_0^2}{8\pi G}$, where $H_0$ is the Hubble constant
and assuming that no shell crossing occurs so that the total mass inside
\begin{figure}
\psfig{file=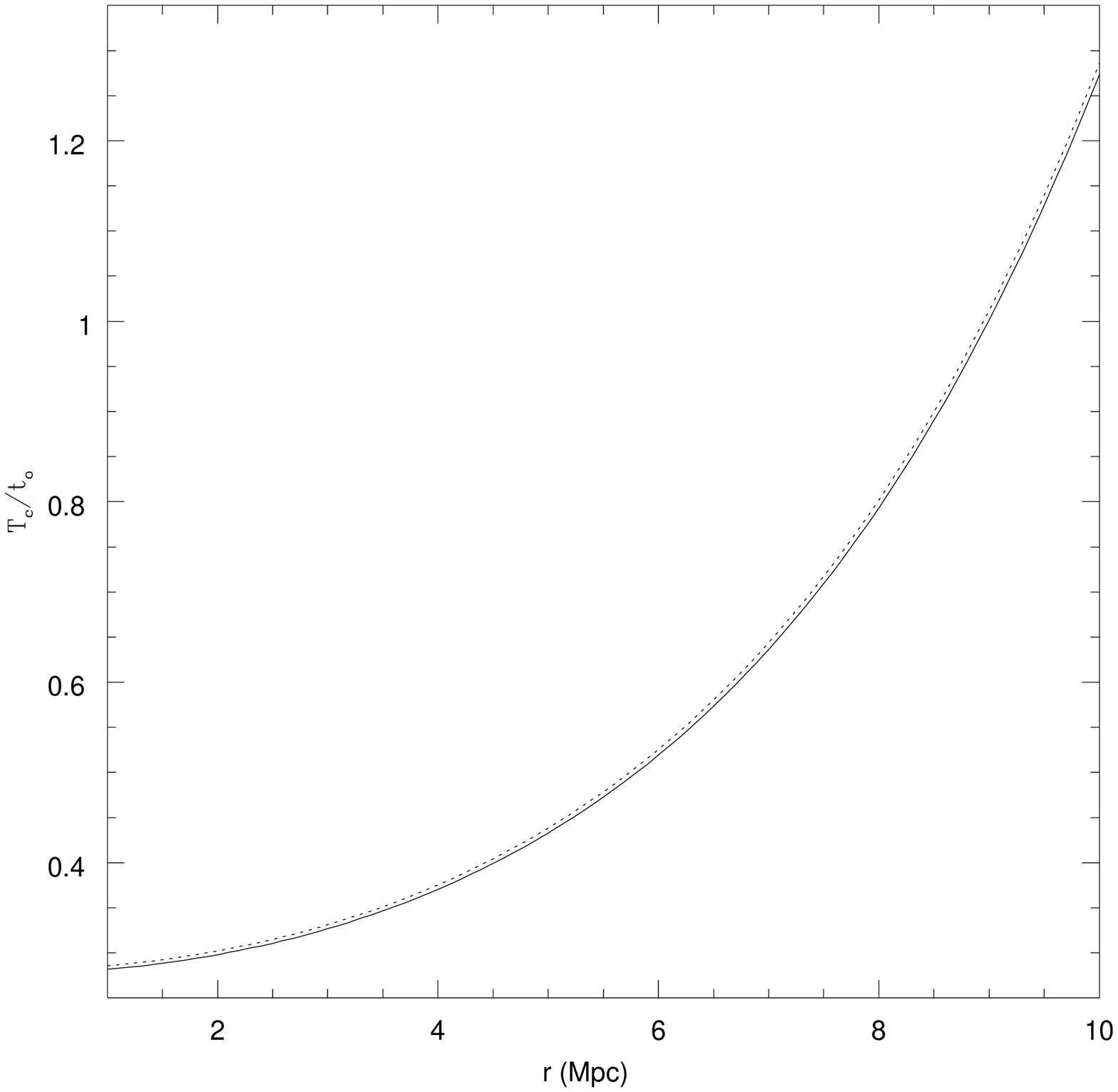,width=11cm,height=11cm}
\caption[]{The time of collapse of a shell of matter in units of the age of the
universe $t_{o}$ for $\nu=4$ (dotted line) compared with Gunn \& Gott's
model (solid line).}
\end{figure}
each shell remains constant, that is:
\begin{equation}
\rho (r_i,t)=\frac{\rho _i(r_i,t)}{a^3(r_i,t)}
\end{equation}
the Eq. (\ref{eq:coll}) may be written as: 
\begin{equation}
\frac{d^2a}{dt^2}=-\frac{H^2(1+\overline{\delta })}{2a^2}+\frac{4G^2L^2}{%
H^4(1+\overline{\delta })^2r_i^{10}a^3} -\eta \frac{da}{dt} \label{eq:sec}
\end{equation}
The equation (\ref{eq:sec}) 
may be
solved using the initial conditions: $(\frac{da}{dt})=0$, $a=a_{max}\simeq 1/%
\overline{\delta }$ and using the function $h(r,\nu )=L(r,\nu )/M_{sh}$
found in \S 2 to obtain a(t) and the time of collapse, $T_c(r,\nu )$. \\ 
In Fig. 2 we show the effects of non-radial motions and dynamical friction
separately. As displayed non-radial motions have a larger effect on the
collapse delay with respect to dynamical friction.
In Figs. 3 $%
\div $ 5 we compare the results for the time of collapse, $T_c$, for $\nu
=2,3,4$ with the time of collapse of the classical GG spherical model: 
\begin{equation}
T_{c0}(r,\nu )=\frac \pi {H_i}[\overline{\delta }(r,\nu )]^{-3/2}
\end{equation}
As shown the presence of non-radial motions produces an increase in the time
of collapse of a spherical shell. The collapse delay is larger for low value
of $\nu $ and becomes negligible for $\nu \geq 3$. This result is in
agreement with the angular momentum-density anticorrelation effect: density
peaks having low value of $\nu $ acquire a larger angular momentum than high 
$\nu $ peaks and consequently the collapse is more delayed with respect to
high $\nu $ peaks. Given $T_c(r,\nu )$ we also calculated the total mass
gravitationally bound to the final non-linear configuration. There are at
least two criteria to establish the bound region to a perturbation $\delta
(r)$: a statistical one (Ryden 1988b), and a dynamical one (Hoffman \&
Shaham 1985). The dynamical criterion, that we have used, supposes that the
binding radius is given by the condition that a mass shell collapse in a
time, $T_c$, smaller than the age of the universe $t_0$: 
\begin{equation}
T_c(r,\nu )\leq t_0
\end{equation}
We calculated the time of collapse of GG spherical model, $T_{c0}(r,\nu )$,
using the density profiles given in Ryden \& Gunn  (1987) for $1.7<\nu <4$ and
then repeated the calculation taking into account non-radial motions
obtaining $T_c(r,\nu )$. Then we calculated the binding radius, $r_{bo}(\nu )
$, for a GG model solving $T_{c0}(r,\nu )\leq t_0$ for $r$ and for several
value of $\nu $, while we calculated the binding radius of the model that
takes into account non-radial motions, $r_b(\nu )$, repeating the
calculation, this time with $T_c(r,\nu )\leq t_0$. 
\begin{figure}
\psfig{file=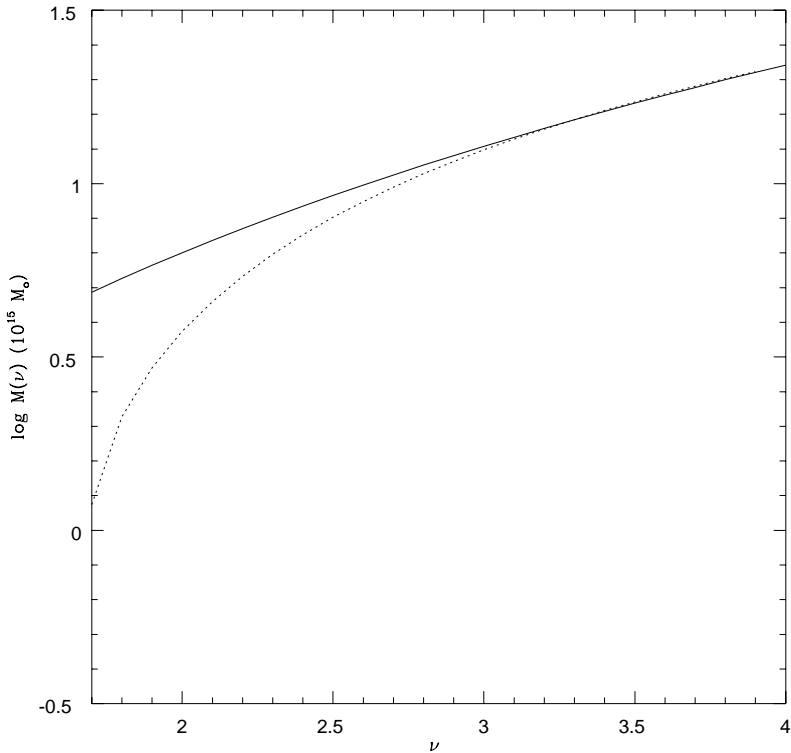,width=11cm,height=11cm}
\caption[]{The mass accreted by a collapsed perturbation, in units of
$10^{15}M_{\odot}$, taking into account non-radial motions 
and dynamical friction effect (dotted line)  
compared to Gunn \& Gott's mass (solid line).}
\end{figure}
We found a relation between $\nu $ and the mass of the cluster using the
equation: $M=\frac{4\pi }3r_b^3\rho _b$.\\ In fig. 6 we compare the peak
mass obtained from GG model, using Hoffman \& Shaham's (1985) criterion,
with that obtained from the model taking into account non-radial motions. As
shown for high values of $\nu $ ($\nu \geq 3$) the two models give the same
result for the mass while for $\nu <3$ the effect of non-radial motions
produces less bound mass with respect to GG model. 
decreases the effect of non-radial
motions produces a decrease in the bound mass.

{\bf 5. Tidal field and the selection function}

Following BBKS we define a selection function 
$t(\nu /\nu _t)$ which gives the probability that a density
peak forms an object, while the threshold level, $\nu _t$, is defined so
that the probability that a peak form an object is 1/2 when $\nu =\nu _t$. 
The selection function introduced by BBKS (Eq. 4.13), is an empirical one

and depends on two parameters: the threshold $\nu _t$ and the shape
parameter $q$: 
\begin{equation}
t(\nu /\nu _t)=\frac{(\nu /\nu _t)^q}{1+(\nu /\nu _t)^q}
\end{equation}
If $q\rightarrow \infty $ this selection function is a Heaviside function $%
\vartheta (\nu -\nu _t)$ so that peaks with $\nu >\nu _t$ have a probability
equal to 100\% to form objects while peaks with $\nu \leq \nu _t$ do not
form objects. If $q$ has a finite value sub-$\nu _t$ peaks are selected with
non-zero probability. Using the given selection function the cumulative
number density of peaks higher than $\nu $ is given, according to BBKS, by: 
\begin{equation}
n_{pk}=\int_\nu ^\infty t(\nu /\nu _t)N_{pk}(\nu )d\nu 
\end{equation}
where $N_{pk}(\nu )$ is the comoving peak density (see BBKS Eq.~4.3). A form
of the selection function, physically motivated, can be obtained following
the argument given in CAD. In this last paper the selection function is
defined as: 
\begin{equation}
t(\nu )=\int_{\delta _c}^\infty p\left[ \overline{\delta },\langle \overline{%
\delta }\rangle (r_{Mt},\nu ),\sigma _{\overline{\delta }}(r_{Mt},\nu
)\right] d\delta   \label{eq:sel}
\end{equation}
where the function 
\begin{equation}
p\left[ \overline{\delta },\langle \overline{\delta }\rangle (r)\right]
=\frac 1{\sqrt{2\pi }\sigma _{\overline{\delta }}}\exp \left( -\frac{|%
\overline{\delta }-\langle \overline{\delta }\rangle (r)|^2}{2\sigma _{%
\overline{\delta }}^2}\right)   \label{eq:gau}
\end{equation}
gives the probability that the peak overdensity is different from the
average, in a Gaussian density field. The selection function depends on $\nu 
$ through the dependence of $\overline{\delta }(r)$ from $\nu $. As
displayed the integrand is evaluated at a radius $r_{Mt}$ which is the
typical radius of the object we are selecting. Moreover the selection
function $t(\nu )$ depends on the critical overdensity threshold for the
collapse, $\delta _c$, which is not constant as in a spherical model (due to
the presence, in our analysis, of non-radial motions and dynamical friction 
that delay the collapse
of the proto-cluster) but it depends on $\nu $. 
An
analityc determination of $\delta _c(\nu )$ can be obtained following a
technique similar to that used by Bartlett \& Silk (1993). Using Eq.~(\ref
{eq:sec}) it is possible to obtain the value of the expansion parameter of
the turn around epoch, $a_{max}$, which is characterized by the condition $%
\frac{da}{dt}=0$. Using the relation between $v$ and $\delta _i$, in linear
theory (Peebles 1980), we find 
\begin{equation}
\delta _c(\nu )=\delta _{co}\left[ 1+\frac{\lambda_{o}}{1-\mu(\delta)}+ 
\frac{8G^2}{\Omega _o^3H_o^6r_i^{10}%
\overline{\delta }(1+ 
\overline{\delta })^2}\int \frac{L^2da}{a^3}\right] 
\end{equation}
where $\delta _{co}=1.68$ is the critical threshold for GG model and $
\lambda_{o}$ and $ \mu(\delta)$ are given in Colafrancesco, Antonuccio \& Del
Popolo (1995) (Eq. 5, 6). 
In Fig. 7 we show the overdensity threshold in function of $\nu $. As shown, 
$\delta _c(\nu )$ decreases with increasing $\nu $. When $\nu $ $>3$ the
threshold assume the typical value of the spherical model. This means,
according to the cooperative galaxy formation theory, (Bower et al. 1993)
that structures form more easily if there are other structures nearby, i.e.
the threshold level is a decreasing function of the mean mass density. Known 
$\delta _c(\nu )$ and chosen a spectrum, the selection function is
immediately obtainable through Eq. (\ref{eq:sel}) and Eq. (\ref{eq:gau}).
The result of the calculation, plotted in fig. 8, for two values of the
filtering radius, ($R_f=2$, $3$ $h^{-1}Mpc$), shows that the selection
function, as expected, differs from an Heaviside function (sharp threshold).
The value of $\nu $ at which the selection function reaches the value 1 ($%
t(\nu )\simeq 1$) increases for increasing values of the filtering radius, $%
R_f$. This is due to the smoothing effect of the filtering process. The
effect of non-radial motions is, firstly, that of shifting $t(\nu )$ towards
higher values of $\nu $, and, secondly, that of making it steeper. The
selection function is also different from that used by BBKS (tab. 3a).
Finally it is interesting to note that the selection function defined by Eq.
(\ref{eq:sel}) and Eq. (\ref{eq:gau}) is totally general, it does not depend
on the presence or absence of non-radial motions. The latter influence the
selection function form through the changement of $\delta _c$ induced by
non-radial motions itself.

{\bf 6. The bias coefficient}
One way of defining the bias
coefficient of a class of objects is that given by (BBKS): 
\begin{equation}
b(R_f)= \frac{<\widetilde{\nu}> }{\sigma_o}+1
\end{equation}
where $\langle \widetilde{\nu} \rangle$ is: 
\begin{equation}
<\widetilde{\nu}> = \int_0^\infty \left[ \nu -\frac{\gamma \theta }{%
1-\gamma ^2}\right] t(\frac{\nu}{\nu_{t}}) N_{pk}(\nu ) d\nu  \label{eq:nu}
\end{equation}
from Eq. (\ref{eq:nu}) it is clear that the bias parameter can be calculated
once a spectrum, $P (k)$, is fixed. The bias parameter depends on the shape
and normalization of the power spectrum. A larger value is obtained for
spectra with more power on large scale (Kauffmann et al. 1996). In this
calculation we continue to use the standard CDM spectrum ($\Omega_0 = 1$, $h
= 0.5$) normalized imposing that the rms density fluctuations in a sphere of
radius $8 h^{-1}Mpc$ is the same as that observed in galaxy counts, i.e. $%
\sigma _8=\sigma (8h^{-1}Mpc)=1$. The calculations have been performed for
three different values of the filtering radius ($R_f=2,$ $3$, $4$ $h^{-1}Mpc$%
). The result of the calculation is plotted in table 1. As shown, the value
of the bias parameter tends to increase with $R_f$ due the filter effect of $%
t(\nu)$. As shown $t(\nu)$ acts as a filter, increasing the filtering
radius, $R_{f}$, the value of $\nu$ at which $t(\nu) \simeq 1$ increases .
In other words when $R_{f}$ increases $t(\nu)$ selects density peaks of
larger height. The reason of this behavior must be searched in the smoothing
effect that the increasing of the filtering radius produces on density
peaks. When $R_{f}$ is increased the density field is smoothed and $t(\nu)$
has to shift towards higher value of $\nu$ in order to select a class of
object of fixed mass, $M$.

\begin{center}
\begin{table}[ht]
\medskip
\begin{tabular}{|l|l|}
\multicolumn{2}{c}{\bf Bias} \\ \hline
$R_f (h^{-1}Mpc) $ & $b$ \\ \hline
2 & $1.6 $ \\ 
3 & $1.93 $ \\ 
4 & $2.25 $ \\ \hline
\end{tabular}
\medskip
\caption{Values of the coefficient of bias for different values of the
filtering radius.}
\end{table}
\end{center}

\begin{figure}[ht]
\psfig{file=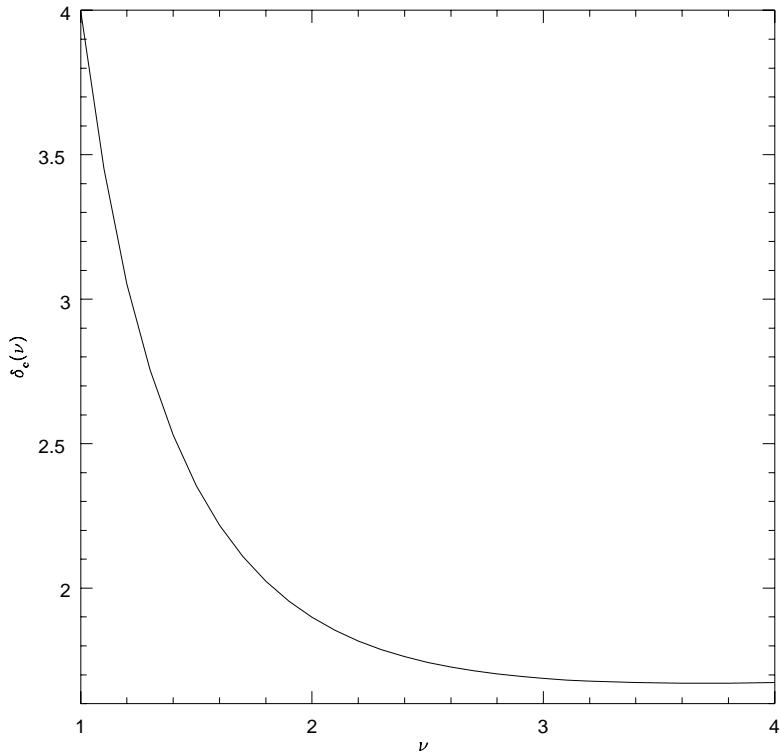,width=11cm,height=11cm}
\caption[]{The critical threshold, $\delta_{c}(\nu)$ versus $\nu$}
\end{figure}

\begin{figure}[ht]
\psfig{file=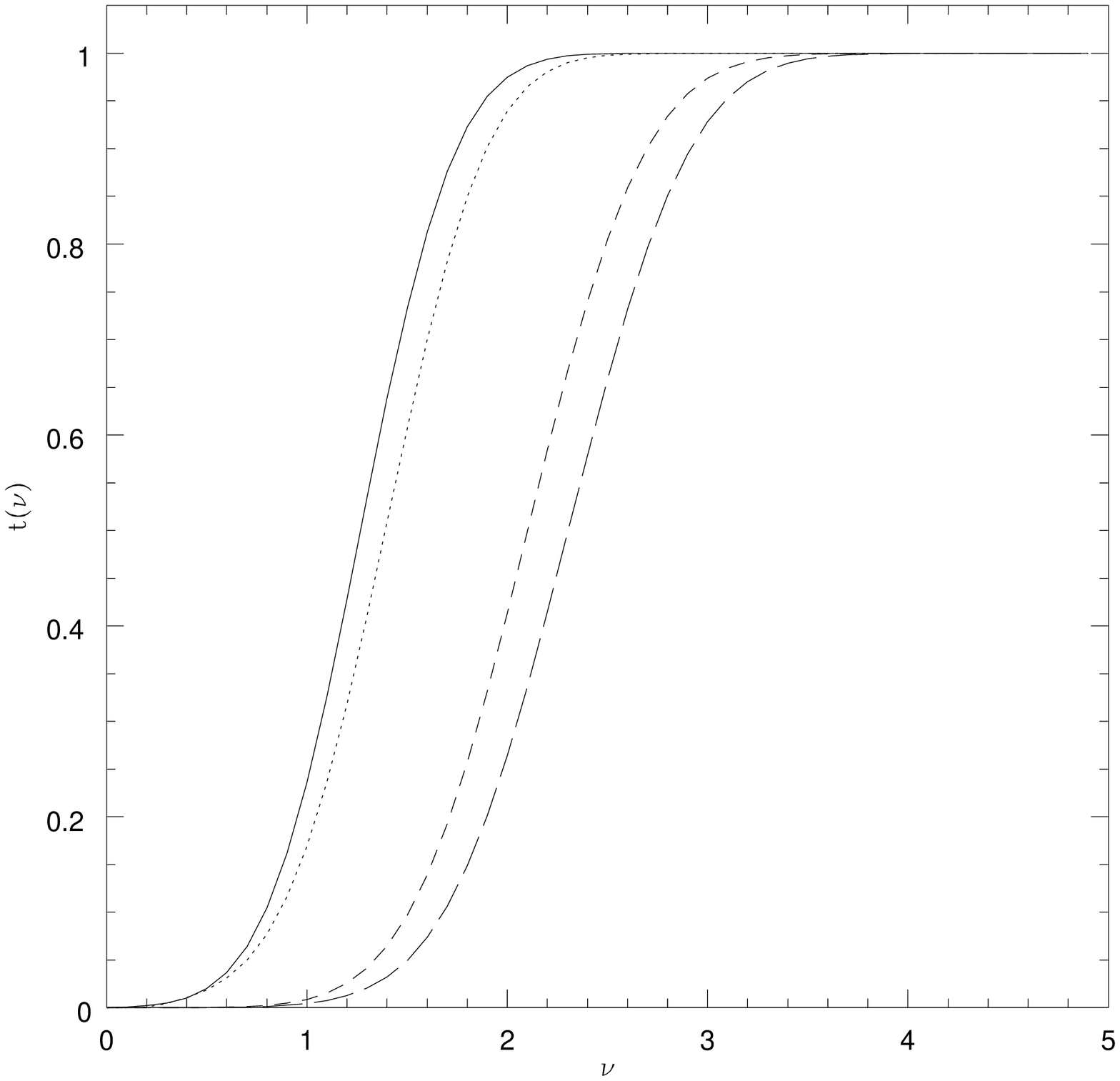,width=11cm,height=11cm}
\caption[]{The selection function, $t(\nu)$, for $R_{f}= 3h^{-1}Mpc$
($\delta_{c}=1.68$, solid line; $\delta_{c}$ function of $\nu$, dotted line) 
and for $4h^{-1}Mpc$ ($\delta_{c}=1.68$, short dashed line;
$\delta_{c}$ function of $\nu$, long dashed line).}
\end{figure}
{\bf 7. Conclusions}

In this paper we have studied the role of non-radial motions and dynamical
friction on the collapse
of density peaks solving numerically the equations of motion of a shell of
barionic matter falling into the central regions of a cluster of galaxies.
We have shown that non-radial motions and dynamical friction 
produce a delay in the collapse of
density peaks having low value of $\nu$ while the collapse of density peaks
having $\nu > 3$ is not influenced. A first consequence of this effect is a
reduction of the mass bound to collapsed perturbations and a raising of the
critical threshold, $\delta_{c}$, which now is larger than that of the
top-hat spherical model and depends on $\nu$. This means that shells of
matter of low density have to be subjected to a larger gravitational
potential, with respect to the homogeneous GG model, in order to collapse.
The delay in the proto-structures collapse gives rise to a dynamical bias
similar to that described in CAD whose bias parameter may be obtained once a
proper selection function is defined. The selection function found is not a
pure Heaviside function and is different from that used by BBKS to study the
statistical properties of clusters of galaxies. Its shape depends on the
effect of non-radial motions and dynamical friction 
through its dependence on $\delta_{c}(\nu)$.
The function $t(\nu)$ selects density peaks higher and higher with
increasing value of $R_{f}$ due to the smoothing effect of the density field
produced by the filtering procedure. Using this selection function and BBKS
prescriptions we have calculated the coefficient of bias, $b$. On clusters
scales for $R_{f} = 4h^{-1}Mpc$ we found a value of $b= 2.25$ comparable
with that obtained from the mean mass-to-light ratio of clusters, APM
survey, or from N-body simulations combined with hydrodynamical models
(Frenk et al. 1990). Besides, the value of the coefficient of biasing $b$
that we have calculated is comparable with the values of $b$ given by
Kauffmann et al. 1996. This means that non-radial motions and dynamical
friction play a significant
role in determining the bias level.


\begin{flushleft}
{\it Acknowledgements}

\end{flushleft}

\end{document}